# Digital image correlation and finite element modelling as a method to determine mechanical properties of human soft tissue *in vivo*

Kevin M. Moerman[a], Cathy A. Holt[b], Sam L. Evans[b], Ciaran K. Simms[a]

[a] Centre for Bioengineering Trinity College Dublin, Ireland
[b] School of Engineering, Cardiff University

*Corresponding author:*
Kevin Mattheus Moerman
E-mail: kevin.moerman@gmail.com

## ABSTRACT

The mechanical properties of human soft tissue are crucial for impact biomechanics, rehabilitation engineering and surgical simulation. Validation of these constitutive models using human data remains challenging and often requires the use of non-invasive imaging and inverse finite element (FE) analysis. Post processing data from imaging methods such as tagged magnetic resonance imaging (MRI) can be challenging. Digital Image Correlation (DIC) however is a relatively straightforward imaging method and thus the goal of this study was to assess the use of DIC in combination with FE modelling to determine the bulk material properties of human soft tissue. Indentation experiments were performed on a silicone gel soft tissue phantom. A two camera DIC setup was then used to record the 3D surface deformation. The experiment was then simulated using a FE model. The gel was modelled as Neo-Hookean hyperelastic and the material parameters were determined by minimising the error between the experimental and FE data. The iterative FE analysis determined material parameters ($\mu$=1.80 kPa, $\kappa$=2999 kPa) which were in close agreement with parameters derived independently from regression to uniaxial compression tests ($\mu$=1.71 kPa, $\kappa$=2857 kPa). Furthermore, the FE model was capable of reproducing the experimental indentor force as well as the surface deformation found ($R^2$=0.81). It was therefore concluded that a two camera DIC configuration combined with FE modelling can be used to determine the mechanical properties of materials that can be represented using hyperelastic Neo-Hookean constitutive laws.

# 1. INTRODUCTION

Knowledge of the mechanical properties of muscle tissue in compression is crucial for modelling in impact biomechanics (Forbes et al. 2005), rehabilitation engineering (Linder-Ganz et al. 2007), and surgical simulation (Famaey and Sloten 2008; Guccione et al. 2001). Compression tests on fresh *in vitro* porcine samples have shown that skeletal muscle tissue in compression is nonlinear elastic, anisotropic and viscoelastic, and a constitutive model has been proposed (Loocke Van et al. 2006; Loocke Van et al. 2008). However, translation to *in vivo* human tissue presents significant technical difficulties. For example, indentation tests have been performed on skeletal muscle (Gefen et al. 2005; Palevski et al. 2006) but these authors have considered muscle tissue as isotropic and linear in elastic and viscoelastic properties. Some authors have used imaging techniques combined with finite element modelling optimised to reproduce experimental boundary conditions. For instance, using MRI, e.g. for skin (Tada et al. 2006), the heart (Walker et al. 2008) and recently also for skeletal muscle (Ceelen et al. 2008). A more straightforward imaging method, digital image correlation (DIC), has also been used to study the mechanical properties of biological soft tissues e.g. using 2D DIC: on the human tympanic membrane (Cheng et al. 2007), sheep bone callus (Thompson et al. 2007), human cervical tissue (Myers et al. 2008) and recently also using 3D DIC: for the bovine cornea (Boyce et al. 2008) and mouse arterial tissue (Sutton et al. 2008). In all these studies the analysis is limited to planar tissue and/or superficial properties of excised tissue samples and thin tissue layers. The potential of using the surface measurements of 3D DIC to assess mechanical states throughout the bulk of a tissue has been suggested (Spencer et al. 2008) but not yet attempted. This Short Communication assesses, for the first time, the use of 3D DIC and inverse FE analysis to non-invasively determine the bulk material properties of soft tissue which could be applied in vivo.

# 2. METHODS

DIC is an optical method which uses tracking and image registration to measure high resolution 3D deformation. The technique relies on tracking of unique features (such as speckles, Figure 1A) within small image subsets (Figure 1B) which can be imaged from multiple camera angles. DIC can be combined with an iterative FE procedure to optimise the parameters of a constitutive model. To verify this method, indentation tests were performed on a silicone gel (SYLGARD® 527, Dow Corning, MI, USA) phantom (Figure 1A). The material parameters for a hyperelastic (Neo-Hookean) material model of the gel were determined by regression (using Prism 4.0, GraphPad Software Inc.) of the model against uniaxial compression tests up to 50% strain on cubic samples (~10mm). Equation 1 shows the strain energy formulation ($\Psi$) for this material law as a function of the modified principal stretches ($\overline{\lambda}_i$) and the Jacobian (*J*).

$$\Psi(\overline{\lambda}_1, \overline{\lambda}_2, \overline{\lambda}_3, J) = \frac{\mu}{2}\left((\overline{\lambda}_1)^2 + (\overline{\lambda}_2)^2 + (\overline{\lambda}_3)^2 - 3\right) + \frac{\kappa}{2}(J-1)^2 \qquad 1$$

$$\overline{\lambda}_i = J^{-\frac{1}{3}}\lambda_i \qquad J = \lambda_1\lambda_2\lambda_3$$

The parameters *μ* and *κ* define material stiffness and the Poisson's ratio (Equation 2).

$$v = \frac{3\left(\frac{K_0}{\mu_o}\right) - 2}{6\left(\frac{K_0}{\mu_0}\right) + 2} \qquad 2$$

The material was assumed nearly incompressible and thus in the fitting procedure κ was constrained to yield a Poisson's ratio of 0.4997. The parameters found were: $\mu$= 1.71 kPa, $\kappa$=2857 kPa ($R^2$ =0.9979).

The soft tissue phantom was moulded in a cylindrical polypropylene container and black paint speckles (0.2~2mm) were applied to the top surface of the gel. A circular indenter was then used to apply compression (Figure 1A). A two camera DIC configuration (Limess Messtechnik & Software GmbH, Pforzheim, Germany) was used to record the deforming phantom and the movement of the speckle pattern for static compression up to 11.7N load applied to the gel through the indentor. Analysis software VIC-3D (DIC Software, Correlated Solutions, Inc., Columbia, SC, USA) was then used to calculate the deformation of the top surface of the gel and the indentor displacement.

To simulate the compression experiment an axisymmetric FE model of the gel indentation was created using Abaqus 6.7-1 Standard (Dessault Systèmes, Suresnes Cedex, France). The container was assumed rigid (no deformation of the container was observed) and simulated by constraining the gel nodes that would be contact with the container from moving in all directions. The piston was modelled as rigid and the gel phantom was modelled using the Neo-Hookean hyperelastic material model and meshed using 4-node quadrilateral elements. Since the silicone gel is sticky the gel/piston contact was modelled with no slip. The experimental displacement was applied to the piston. The material parameters were then iteratively altered until a good match with both the experimental indentor force and deformation was obtained. The upper region of the FE model is shown in its initial and deformed state in figure 2A and 2B respectively.

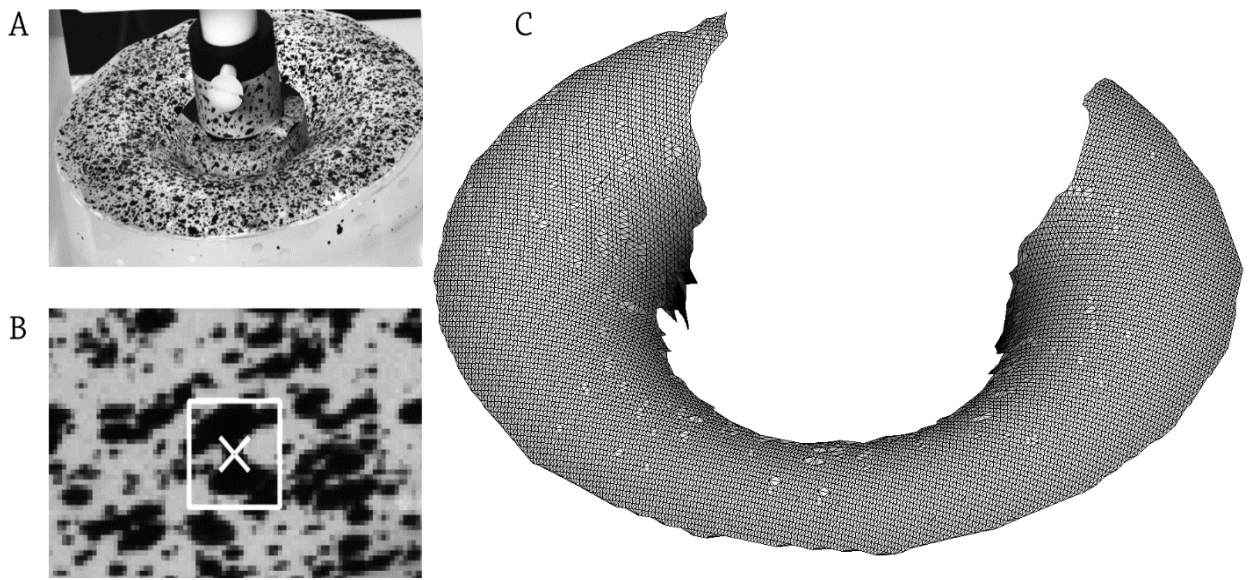

*Figure 1: (A) The silicone gel phantom model and piston,(B) Close-up of speckles showing a subset, (C) A Delaunay triangulation of the tracked data points on the gel surface*

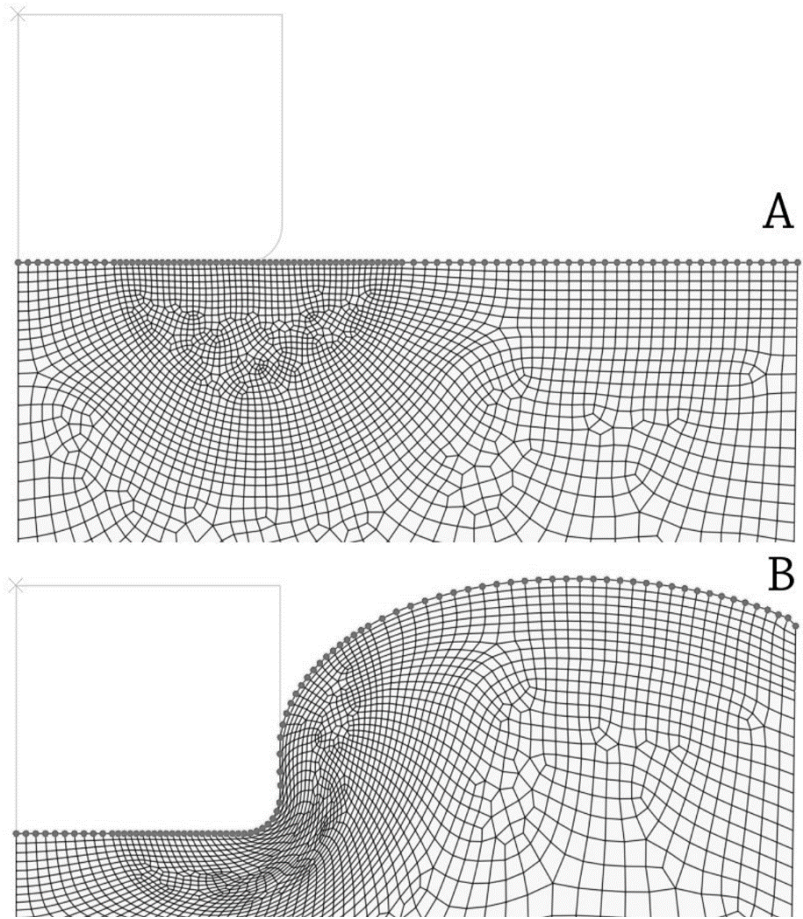
*Figure 2: Top surface region of FE model: (A) Initial FE mesh, (B) deformed FE mesh*

## 3. RESULTS

The experimental and FE results were compared using Matlab 7.4 R2007a (The Mathworks Inc., USA). The best match to the experimental data was achieved using $\mu$=1.80 kPa and $\kappa$=2999 kPa ($v$=0.4997) which are a close match to the parameters derived from uniaxial compression ($\mu$=1.71 kPa, $\kappa$=2857 kPa).

The experimental DIC results are 3D coordinates of points tracked on the top surface of the phantom (Figure 1C). In order to compare the 3D experimental deformation data with the axisymmetric 2D FE deformation data, the 3D data was revolved around the central axes. This effectively "wrapped" all data points around the central axes and mapped them into a single plane. Figure 3A is a top view of the experimental surface at the maximim compression depth (16.75mm). The surface is shaded depending on the error with the FE results. The average error observed is 0.4 mm with the largest errors around the centre where correlation was poorer due to the inward curvature of the gel. A mirrored representation showing the 2D FE results and the wrapped DIC results are shown in Figure 3B. The shape of the top surface in the FE simulation is a good match to the experimental shape ($R^2$=0.81). Figure 4 demonstrates that near incompressibility was a good assumption, as lowering $v$ results in divergence from the experimental deformation ($\mu$ was held constant).

A parametric analysis for the parameter $\mu$ was performed using FE simulations. The resulting indentor force displacement curves are shown in Figure 5. In addition to the experimental indentor force displacement data, curves are shown for the best match (using $\mu$=1.80 kPa) and the $\mu$ derived from regression to uniaxial data ($\mu$ =1.71 kPa). To indicate how altering $\mu$ influences the indentor force response, curves for $\mu$=1.37 kPa and $\mu$=2.05 kPa (20% lower and 20% higher than 1.71 kPa respectively) are also shown.

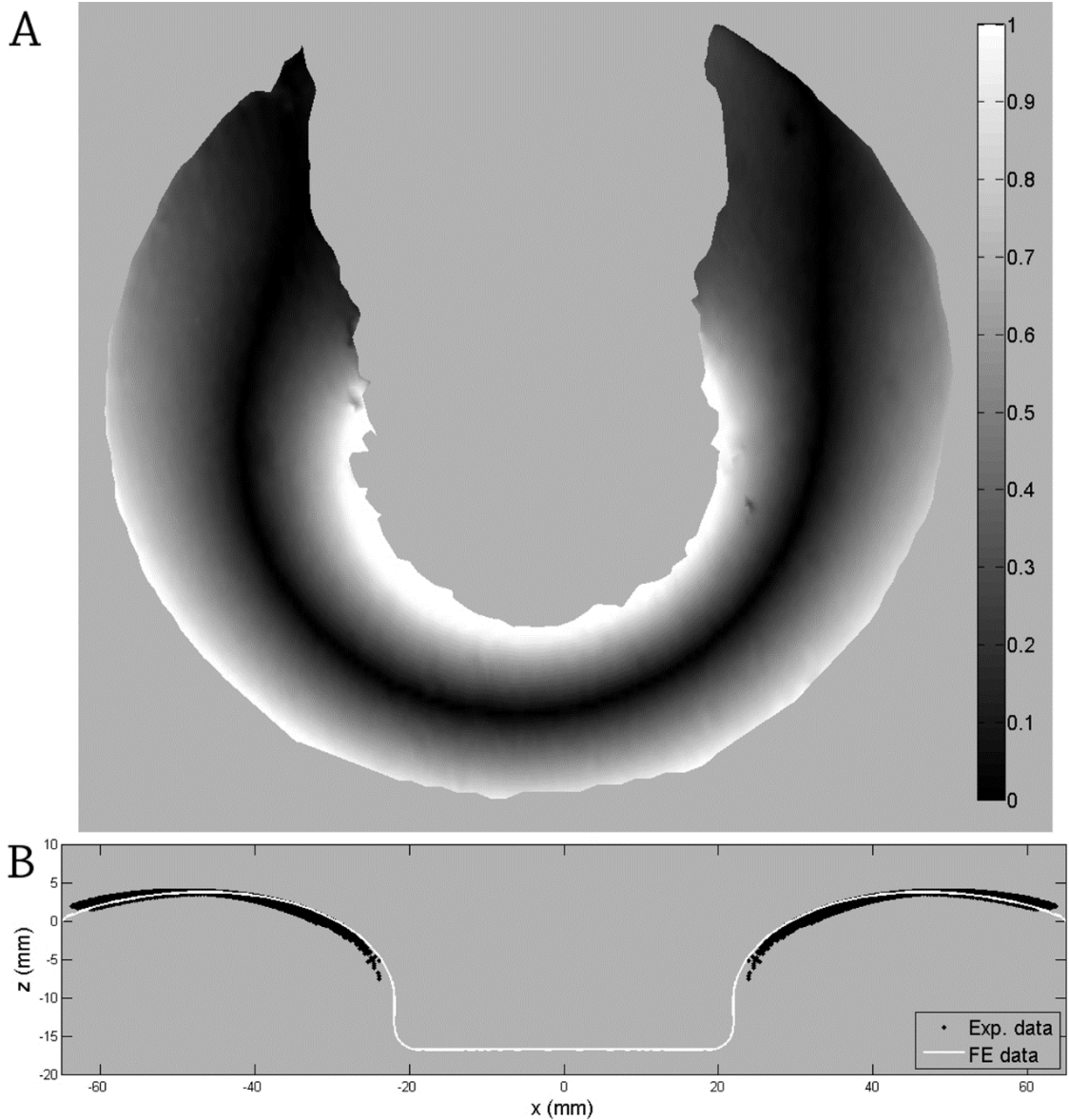

Figure 3: Comparing DIC and FE deformation data. (A) 3D comparison of revolved FE surface with DIC surface. Surface is coloured according to the absolute error (mm). (B) 2D comparison. Average error is 0.4 mm, $R^2$=0.81.

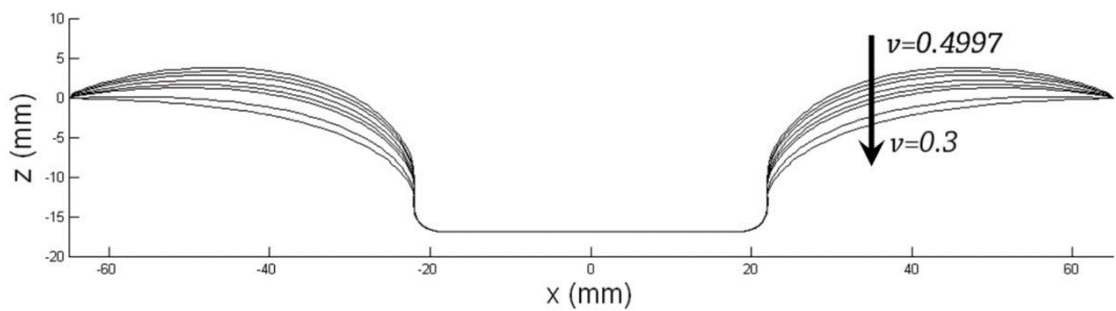

Figure 4: Influence of choice of Poisson's ratio on resulting FE deformation at piston depth of 16.74mm. From top to bottom surface deformation for v=0.4997(best match), v=0.495, v=0.49, v=0.48, v=0.47, v=0.45, v=0.4 and v=0.3

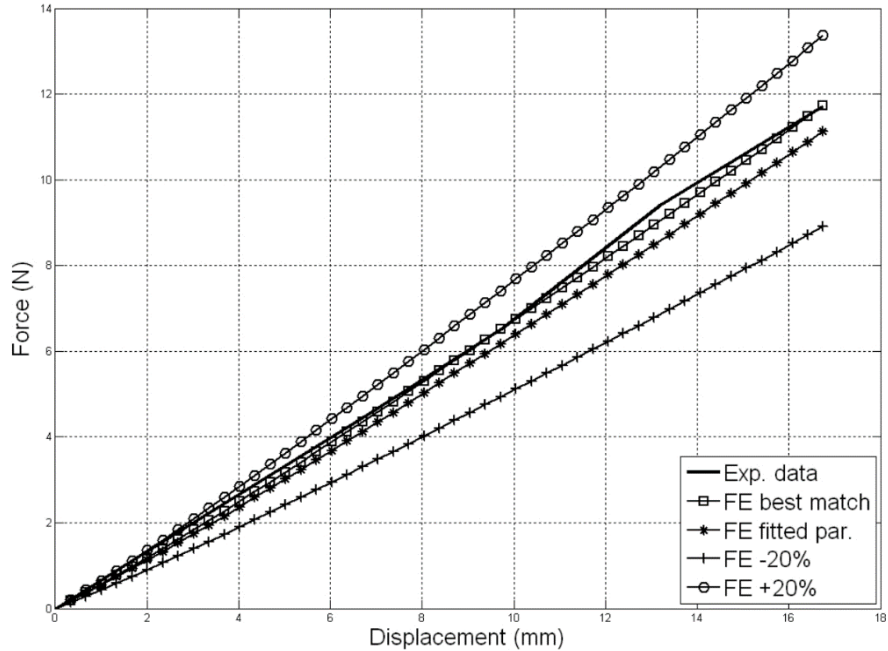

*Figure 5: Indentor force displacement curves. Experimental (solid), FE best match to experimental (solid, square), FE using fitted parameters (solid, star), FE using 20% lower µ (solid, plus), FE using 20% higher µ (solid, circle).*

## 4. DISCUSSION

Figures 3 and 5 demonstrate that the experimental indentor force and shape boundary conditions could be reproduced using $\mu$=1.80 kPa and $\kappa$=2999 kPa (such that $v$=0.4997). The shape of the top surface in the FE simulation mainly depends on the choice of the Poisson's ratio (Figure 4) and the force mainly on the choice of $\mu$ (Figure 5).

The $\mu$ parameter for the best match ($\mu$ =1.80) closely matches the $\mu$ parameter derived from independent uniaxial compression tests ($\mu$ = 1.71). Using $\mu$=1.71 in the FE simulation results in predicted indentor force of 11.14N which is a 95% match to the experimental force (11.7N). However it is likely that the experimental indentor force may have been lower than 11.7N due to friction in the indentation device.

Furthermore, the shape of the deformed finite element surface in Figure 3 is a close match ($R^2$=0.81) to the experimental shape. The small difference may partially be related to the fact that the gel surface meniscus was not modelled (the initial surface in the FE model is flat).

Overall these results demonstrate that when the correct bulk material properties are applied, the FE model has a good capability of reproducing the experimental boundary conditions. It is therefore concluded that a two camera DIC configuration to record 3D surface deformation, in combination with FE modelling, can be used to determine the bulk constitutive parameters of hyperelastic Neo-Hookean materials. The present work is limited to isotropic and elastic materials and thus application to anisotropic and viscoelastic materials requires further research. DIC is more straightforward than other imaging modalities such as MRI and, combined with FE modelling, has the potential to characterize not only the superficial properties but also the underlying bulk constitutive properties for materials with arbitrary shapes undergoing large deformations.

# ACKNOWLEDGEMENTS

Funding: Royal Society (incoming short visit 2007/R3) and Research Frontiers Grant (06/RF/ENM076) awarded by Science foundation Ireland.